\documentclass[aps,pra,twocolumn,floatfix,superscriptaddress]{revtex4-1}
\usepackage{amsmath}
\usepackage{graphicx}
\usepackage{color}
\usepackage{bm}
\usepackage{bigints}
\usepackage[colorlinks]{hyperref}
	\hypersetup{citecolor=red,urlcolor=blue}
\usepackage{blkarray}
\usepackage{tikz}
\usetikzlibrary{shapes,arrows,automata}
\usetikzlibrary{patterns}
\usetikzlibrary{intersections}

\newcommand{\be}{\begin{eqnarray}}
\newcommand{\ee}{\end{eqnarray}}

\newcommand{\beq}{\begin{equation}}
\newcommand{\eeq}{\end{equation}}
\newcommand{\beqa}{\begin{eqnarray}}
\newcommand{\eeqa}{\end{eqnarray}}

\newcommand{\degr}{^{\circ}}

\newcommand{\ctamop}{\affiliation{Centre for Theoretical Atomic, Molecular and Optical Physics, School of Mathematics and Physics, Queen's University Belfast,
\\
University Road, Belfast, BT7 1NN, Northern Ireland, UK}}

\newcommand{\oumk}{\affiliation{Department of Physical Sciences, The Open University, Walton Hall, MK7 6AA Milton Keynes, UK}}

\newcommand\hlight[1]{}

\newcommand\rhlight[1]{}

\newcommand\ghlight[1]{}

\begin{document}

\title{Electron correlation and short-range dynamics in attosecond angular streaking}

\author{G. S. J. Armstrong}
\ctamop
\email[]{gregory.armstrong@qub.ac.uk}
\author{D. D. A. Clarke}
\ctamop
\author{J. Benda} 
\oumk
\author{A. C. Brown}
\ctamop
\author{H. W. van der Hart}
\ctamop

\date{\today}
\begin{abstract}
We employ the $R$-matrix with time-dependence method to study attosecond angular streaking of F$^-$. Using this negative ion, free of long-range Coulomb interactions, we elucidate the role of short-range electron correlation effects in an attoclock scheme. Through solution of the multielectron time-dependent Schr\"{o}dinger equation, we aim to bridge the gap between experiments using multielectron targets, and one-electron theoretical approaches. We observe significant negative offset angles in the photoelectron momentum distributions, despite the short-range nature of the binding potential. We show that the offset angle is sensitive to the atomic structure description of the residual F atom.  We also investigate the response of co- and counter-rotating electrons, and observe an angular separation in their emission.
\end{abstract}


\maketitle

Attosecond angular streaking \cite{eckle2008}, or the attoclock, was developed with the aim of interrogating the electron tunneling process on the attosecond timescale. In the attoclock scheme, a few-cycle, intense, near-circularly-polarized laser pulse modifies the binding potential of the target atom or ion to form a rotating barrier, through which an electron may tunnel. The electron then emerges from the barrier, most probably at a time when the laser field is maximal. Use of an ultrashort pulse serves to localize the ejected-electron wavepacket within an angular interval, and a semiclassical analysis dictates that the the most probable photoelectron momentum should lie along the semi-major axis of the laser vector potential polarization ellipse. However, experiments typically observe that the peak is offset from this direction by some angle, from which a `tunneling time' is inferred.

Experimental investigations of this process have yielded seemingly contradictory conclusions. Early attoclock measurements observed that the most probable emission occurred at an angle to the major axis of the laser polarization ellipse. From the measured angular offsets, tunneling times less than 10 attoseconds were deduced for argon and helium \cite{eckle2008,eckle2008sc,pfeiffer2012}, implying that tunneling occurs effectively instantaneously. Later experiments investigating helium, argon, and krypton targets \cite{landsman2014,camus2017}, favored the interpretation of non-instantaneous tunneling in these systems. Furthermore, using an approach beyond the scope of traditional ultrafast measurements, experimental work has addressed the tunneling of rubidium atoms in an optical lattice, observing tunneling times on the order of microseconds \cite{fortun2016}. 

On the theoretical side, there appears to be increasing consensus. The tunneling process in hydrogen was deemed to be near-instantaneous in a number of theoretical studies \cite{ni2016,torlina2015,bray2018,sainadh2019}, which raised the debate regarding the interpretation of experimental attoclock offset angles \cite{hofmann2019}. However, reduced dimensionality calculations indicate finite tunneling delays \cite{teeny2016}. Controversy also exists regarding the adiabaticity of the tunneling process, with theory and experiment reaching different conclusions in the case of helium \cite{boge2013,ivanov2014,hofmann2014}. Determination of the tunneling adiabaticity has important consequences for laser intensity calibration \cite{klaiber2015}. Further theoretical insights are currently being provided by alternative attoclock schemes, using two-color pulses \cite{han2019}, and tailored fields \cite{eicke2020}. The controversy surrounding interpretation of the attoclock is, at least in part, due to the differing conclusions reached by the wide variety of theoretical methods used for its study. These methods vary in their treatment of adiabaticity, dimensionality, core polarization,
ionized electron motion and multielectron effects. Ultimately, this could be resolved by a theoretical method capable of treating as many of these effects as possible from first principles.


Naturally, a key element of any tunneling process is the potential barrier experienced by the outgoing electron. This has been the focus of most theoretical investigations of the attoclock scheme. The desire to probe the tunneling process in more detail has prompted the application of a number of electron-trajectory  methods, including the analytical $R$-matrix method \cite{torlina2015},  the so-called `Keldysh-Rutherford model' \cite{bray2018}, the classical backpropagation approach \cite{ni2016,ni2018}, and a Bohmian formulation \cite{douguet2018}. Calculations for H \cite{torlina2015,bray2018,sainadh2019} showed that offset angles in momentum distributions were significant when the physical Coulomb potential was retained, but vanished when a Yukawa potential was used. This suggested that offset angles were primarily induced by deflection of the outgoing electron in the Coulomb potential of the residual ion, and less so (or not at all) by a finite time interval spent by the electron tunneling through the potential barrier. Particularly compelling theoretical evidence has been supplied by the Keldysh-Rutherford model of the attoclock \cite{bray2018}, which demonstrated a clear similarity between offset angles and classical Rutherford scattering angles. However, as noted in Ref. \cite{douguet2019}, the use of an adjustable screening length in the Yukawa potential renders its use somewhat arbitrary. Indeed, non-zero offset angles can be realised simply by increasing the screening length beyond unity, and only as the latter tends to infinity are the results for a true Coulomb potential recovered.


Recently, attempts have been made to disentangle Coulomb interactions from the tunneling process in multielectron systems through a judicious choice of target, namely negative ions \cite{douguet2019}. In such systems, the residual neutral and ejected electron interact through a short-range polarization potential. Measurement of the attoclock offset angles for such systems would therefore provide further information on the tunneling process, free from the influence of long-range Coulomb interactions. The study of Ref.\;\cite{douguet2019} investigated detachment from F\(^-\) and Cl\(^-\) exposed to 1500-nm pulses, employing a one-electron approach that accounted for the induced dipole moment of the neutral F residue through a time-dependent, core polarization potential. For these negative ions, angular offsets close to zero degrees in polarization-plane momentum distributions were calculated for both circularly and elliptically polarized pulses.

Despite the apparent consensus of these studies that offset angles are caused by the Coulomb potential and not by tunneling time, theoretical approaches have been largely confined to a single-active-electron (SAE) response. There is clearly a distance between these approaches and the experiments carried out on noble-gas targets, to which such methods are fundamentally unsuited. It is rapidly becoming apparent that this discrepancy may be due to multielectron effects, or interactions at short range in the residual ion. To date, the only exploration of electron correlation in attoclock studies have been carried out for helium \cite{agapi2015,majety2017}. However, it is precisely such effects that we aim to address in this work for a larger, multielectron system.

In this work, we employ the R-matrix with time-dependence (RMT) method \cite{nikolopoulos2008,moore2011,clarke2018,rmtcpc} to study angular streaking from F$^-$. RMT is an {\em ab initio} method that solves the time-dependent Schr\"{o}dinger equation for multielectron atoms, ions and molecules in strong laser fields. The method divides position space into two distinct regions, according to radial distance from the nucleus. An inner region is confined to small distances, and encapsulates the target nucleus. This region contains a truly many-body wave function, that accounts for short-range electron exchange and electron-electron correlation. An outer region extends to large distances from the nucleus, and contains a single, ionized electron that is subject to the interaction with the residual system, as well as the laser field.
Note that by virtue of our fully quantum mechanical treatment, we need not appeal to a classical trajectory description for the detached electron, and thus make no assumptions regarding a characteristic tunnel exit radius or initial momentum.


Our treatment of the F$^{-}$ ionic structure is described in previous work \cite{hassouneh2015,clarke2018,armstrong2019}, and is based on earlier $R$-matrix Floquet calculations for this system \cite{vdh1996,vdh2000}.
To investigate electron correlation in the core, we consider two atomic structure models, one in which the residual F atom is treated at the Hartree-Fock level of detail \cite{clemroe}, and another that uses pseudo-orbitals to construct a set of configurations for subsequent use in a configuration-interaction calculation of the \(1s^2 2s^2 2p^5\ ^2P^o\) state of F. We couple a single electron to the residual neutral, retaining all \(1s^2 2s^2 2p^5 \epsilon l\) channels up to \(l=L_{\rm max}+1\). The calculations used $L_{\rm max} = 79$, yielding 6400 $LM_LS\pi$ symmetries and 9639 channels.

The F$^-$ target interacts with a laser field that is treated classically and within the electric dipole approximation. Since the laser interaction is described in the length gauge \cite{hutchinson2010}, we adopt an appropriate form for the electric field. The electric fields used in this work are derived from the vector potential using
\(\bm{\mathcal E}(t) = -\frac{1}{c}\frac{d}{dt}\bm{\mathcal A}(t)\),
where
\begin{equation}
\bm{\mathcal A}(t) = 
-\frac{c{\cal E}_0}{\omega\sqrt{2}} \sin^4\left(\frac{\omega t}{2N_c}\right) 
\left[
\cos\omega t \;\hat{\bf x} + \sin\omega t \;\hat{\bf y}
\right].
\label{afield}
\end{equation}
Here, \(c\) is the speed of light in vacuum, \({\cal E}_0\) is the peak electric field strength, \(\omega\) is the laser frequency, and \(N_c\) is the number of laser cycles. In this work, we consider laser pulses with a range of peak intensities, carrier wavelengths of 800 nm and 1500 nm, and which ramp on over one laser cycle followed by one cycle of ramp-off, so that \(N_{c} = 2\) in all cases.

We propagate the wave function using an Arnoldi propagator, with a timestep \(\delta t = 0.242\) as, for a total of 60~fs, which represents 20 cycles following termination of the 800-nm pulse. At 1500 nm, the total propagation time is 48~fs, equivalent to around 8 field cycles following termination of the pulse. In all cases, the outer region extends to a distance of 2856$a_0$ (with $a_0$ the Bohr radius), which is sufficient to contain the ejected-electron wave function throughout the time propagation, and a spatial finite-difference grid spacing of \(\delta r = 0.08 a_0\) is used.


Following the time propagation, we calculate the photoelectron momentum distribution in the laser polarization plane. This is achieved by decoupling the photoelectron wave function from that of the residual, neutral F atom, and transforming to momentum space using a Fourier transform. Note that we include a factor of $k^2$ in the distribution to account for the spherical volume element.
\begin{figure}[t]
    \includegraphics[width=\columnwidth,trim={0cm 2cm 0 0}]{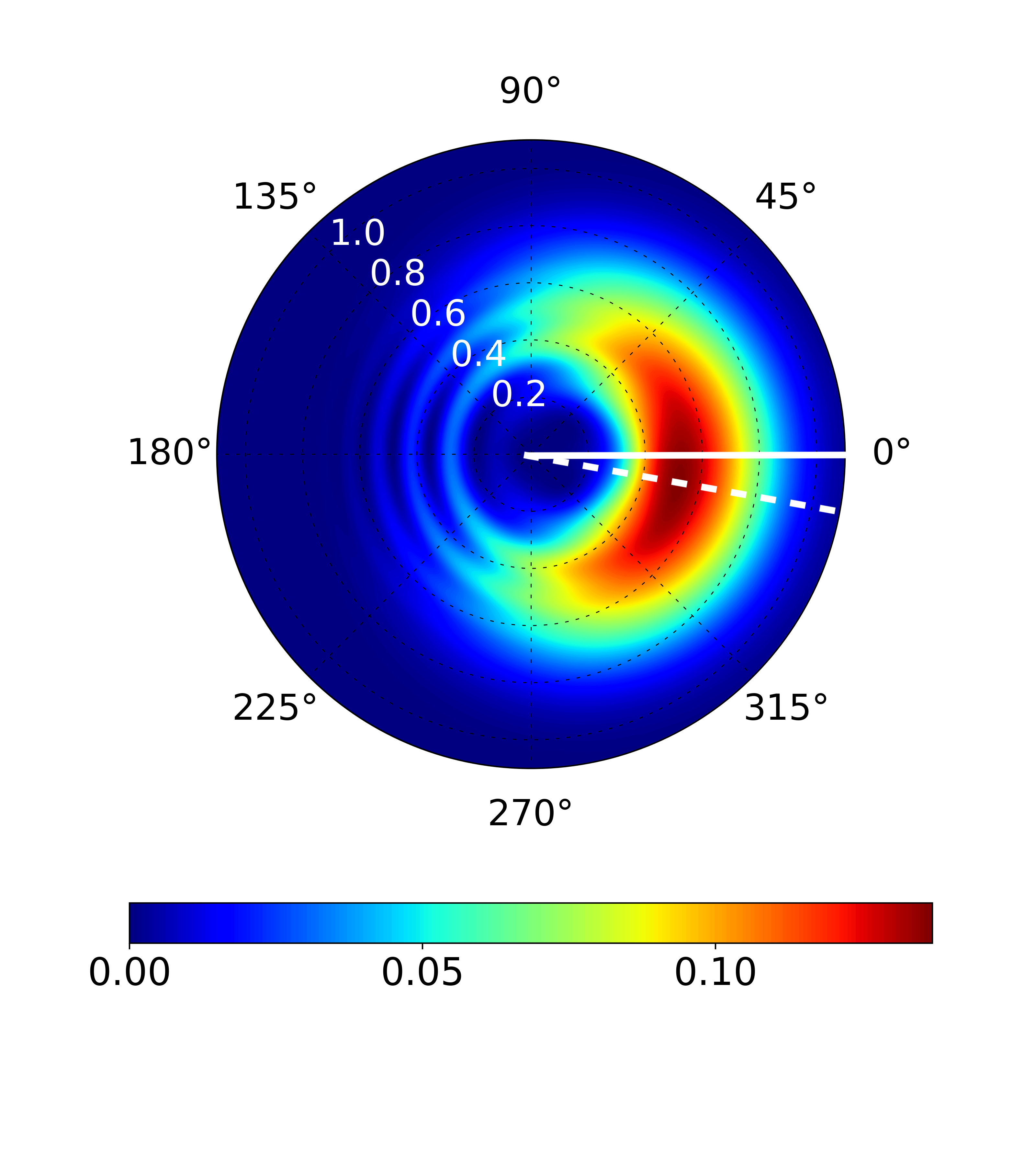}
    \caption{Photoelectron momentum distribution, in the laser polarization plane, arising from F\(^-\) in a  two-cycle, 800-nm, \(5\times10^{13}\) W/cm\(^2\), circularly polarized laser pulse. The offset angle is defined between the maximum of the distribution (dashed line) and the \(x\) axis (solid line).}
    \label{fig:800-51013}
\end{figure}

Figure \ref{fig:800-51013} shows the photoelectron momentum distribution, in the laser polarization plane, for electron detachment from F\(^-\) by an 800-nm pulse of peak intensity \(5\times10^{13}\) W/cm\(^2\).   The distribution is dominated by a broad feature centered close to the horizontal axis. The adiabatic-limit assumption that the electron is detached at the peak (center) of the vector potential with zero initial velocity, implies that the strongest photoelectron yield should lie in the \(\hat{\bf x}\) direction, \(\theta = 0\degr\). However, the peak of the distribution is clearly shifted away from the \(\hat{\bf x}\) direction, and appears at a negative angle. In this case, the peak of the distribution is attained at a rather substantial one of around --12 degrees.


The negative offset angle observed here is, of course, at variance with both the positive offset angles observed in measurements for noble-gas targets, as well as the near-zero offsets observed for short-range potentials in one-electron systems \cite{torlina2015,bray2018,sainadh2019,douguet2019}.
In the absence of a long-range potential, the first obvious source of a negative offset is depletion of the ground state, which is known to be at least partly responsible for calculated offset angles \cite{torlina2015}, without being solely accountable for them.


\begin{figure}[t]
    \centering
    \includegraphics[width=\columnwidth,trim={1cm 1cm 1cm 1cm}]{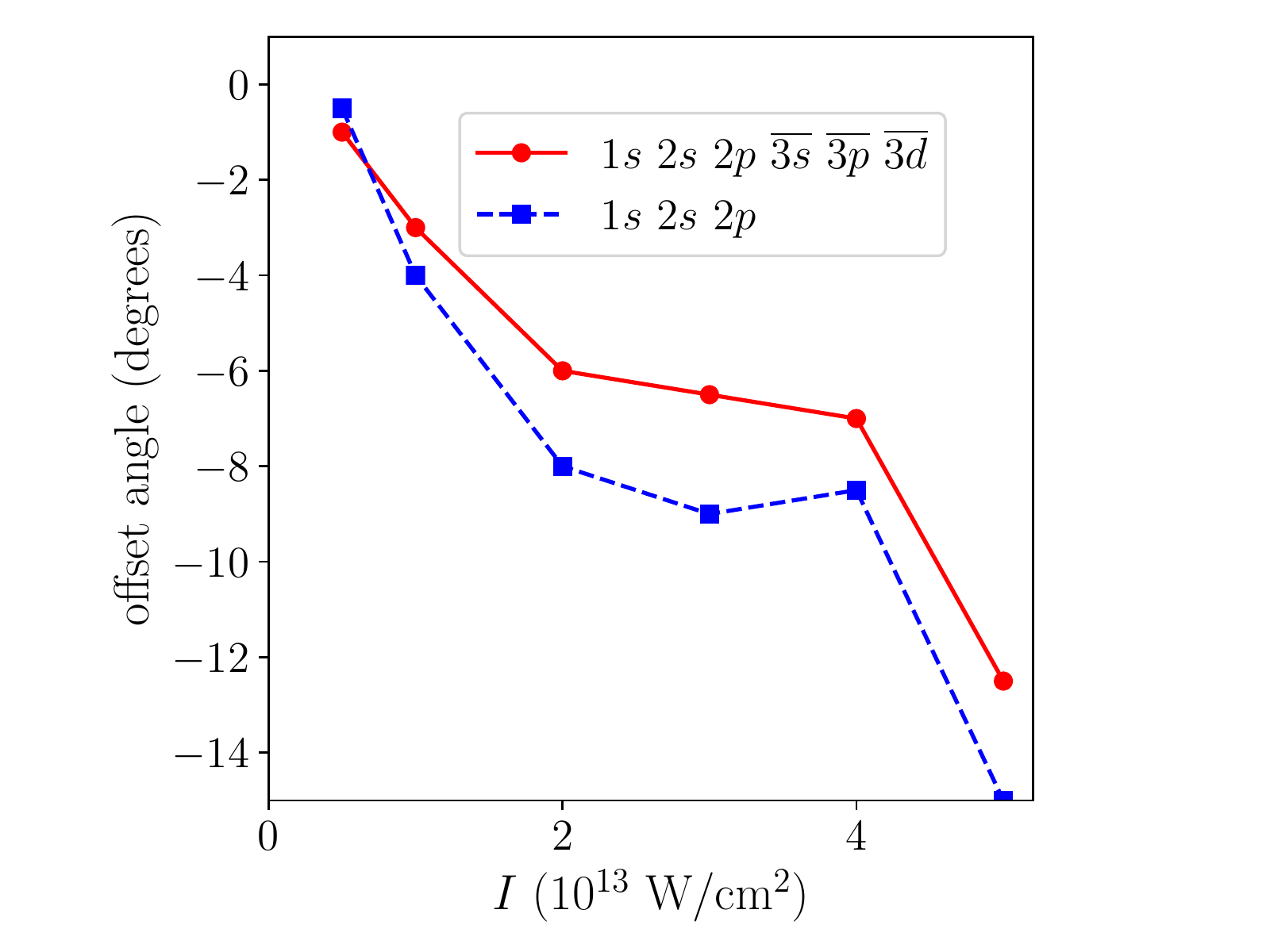}
    \caption{Offset angle as a function of peak laser intensity for F\(^-\), exposed to a two-cycle, 800-nm, circularly polarized laser pulse. Data shown for two atomic structure models, one containing the Hartree-Fock \(1s, 2s\) and \(2p\) orbitals only, and another containing additional \(\overline{3s}, \overline{3p}\) and \(\overline{3d}\) pseudo-orbitals.}
    \label{idep800}
\end{figure}
To address the question of depletion effects under the conditions used here, we calculate the photoelectron momentum distribution over a range of laser intensities, and investigate the variation in offset angle. Fig.\;\ref{idep800} shows the offset angle for peak laser intensities between \(5\times10^{12}\) W/cm\(^2\) and \(5\times10^{13}\) W/cm\(^2\). Over this range of peak intensities, depletion of the ground state varies strongly, from 0.45\% at \(5\times10^{12}\) W/cm\(^2\) to 18\% at \(5\times10^{13}\) W/cm\(^2\). However, the offset angle displays a non-trivial dependence on laser intensity. The offset angle caused by depletion should scale linearly with the ground-state population \cite{torlina2015}, and therefore as a high power of the peak intensity. The variation seen in Fig.\;\ref{idep800} suggests that although depletion undoubtedly has some influence, additional factors affect the offset angle. 

If depletion is not the cause of the offset angle, we must consider the remaining limited set of possibilities. In a negative ion, the main remaining source lies in short-range effects near the nucleus. To investigate the role of short-range interactions with the core, we perform calculations with a second atomic structure model, which uses only the Hartree-Fock orbitals and neglects the pseudo-orbitals. Using this more crude description of the atomic structure, we observe a similar intensity-dependence of the offset angle in Fig.\;\ref{idep800}, but note that the magnitudes of the offsets tend to be lower than those obtained using the higher quality structure model. These results demonstrate that the accuracy of the short-range core potential can have a quantitative bearing on theoretically predicted offset angles, and suggest that electron correlation may indeed play a fundamental role in determining the latter observable for negative ions.


The conclusion reached here regarding electron correlation differs clearly from that demonstrated for helium in Refs.\;\cite{agapi2015,majety2017}, where electron correlation had no significant impact on offset angles. We do not dispute this finding in the case of helium, but rather our results demonstrate that electron correlation should not be assumed negligible for atoms larger than helium. In the case of a negative ion, it may be expected that electron correlation will have a manifested impact on the detachment dynamics, since the outer electron is itself loosely bound by short-range polarization forces. 


A second potential source of negative offset angle is a non-zero initial velocity, with a negative component along the \(y\) direction. 
At the intensities used here, the Keldysh parameter is close to 1, and so the adiabatic-limit assumption of electron detachment with zero initial velocity may not hold. The adiabaticity of the tunneling process has been debated in many studies \cite{pfeiffer2012,boge2013,camus2017}. Here, the choice of a negative ion target allows this aspect of the dynamics to be disentangled from the effects associated with a Coulomb potential. Our numerical method makes no assumptions regarding the initial velocity of the detached electron, and can therefore assess the adiabaticity of the detachment process in a totally unbiased manner.
However, PPT theory predicts that any shifts in the photoelectron momentum due to non-adiabatic effects should occur parallel to the semi-major axis of the laser vector potential (perpendicular to the semi-major axis of the  electric field) \cite{barth2011,hofmann2014}. Given the form of the vector potential in Eq.\;\eqref{afield}, any non-adiabatic tunneling ought to shift the distribution along the $x$ axis. Therefore, it seems unlikely that non-adiabatic dynamics are responsible for the offset angles appearing in our results.

We therefore consider what other short-range effects could be responsible for the observed negative offset angles. In this multielectron system, the ionic fragmentation process involves both the detachment of a single \(2p\) electron, as well as the response of the remaining \(2p^5\) core. Following detachment of a \(2p\) electron, the core \(2p\) electrons naturally adjust. Hartree-Fock calculations estimate that the average orbital radius of a \(2p\) electron in F\(^-\) is 1.25 atomic units, whereas in the neutral F atom, the average radius is 1.08 atomic units \cite{vdh2000}.  This significant retraction of the core electrons is fundamentally a multielectron process, and is captured by RMT but absent in an SAE approach.
The incorporation of pseudo-orbitals in our calculations facilitates a more accurate account of short-range electronic correlations, which in turn may alter the characteristics of the core relaxation dynamics following the photodetachment process. It is plausible that the differing offset angles seen in Fig.\;\ref{idep800} actually manifest the differing treatments of core retraction, suggesting that the latter may have some influence on attoclock measurements for such complex targets.
Note that the effect is made clear through the choice of a negative ion target, free of long-range interactions that are likely to obscure its observation in studies of neutral atoms.

\begin{figure}[t]
$
\begin{array}{cc}
(a) & (b)
\\
        \includegraphics[width=0.5\columnwidth,trim={1cm 2cm 1cm 1cm}]{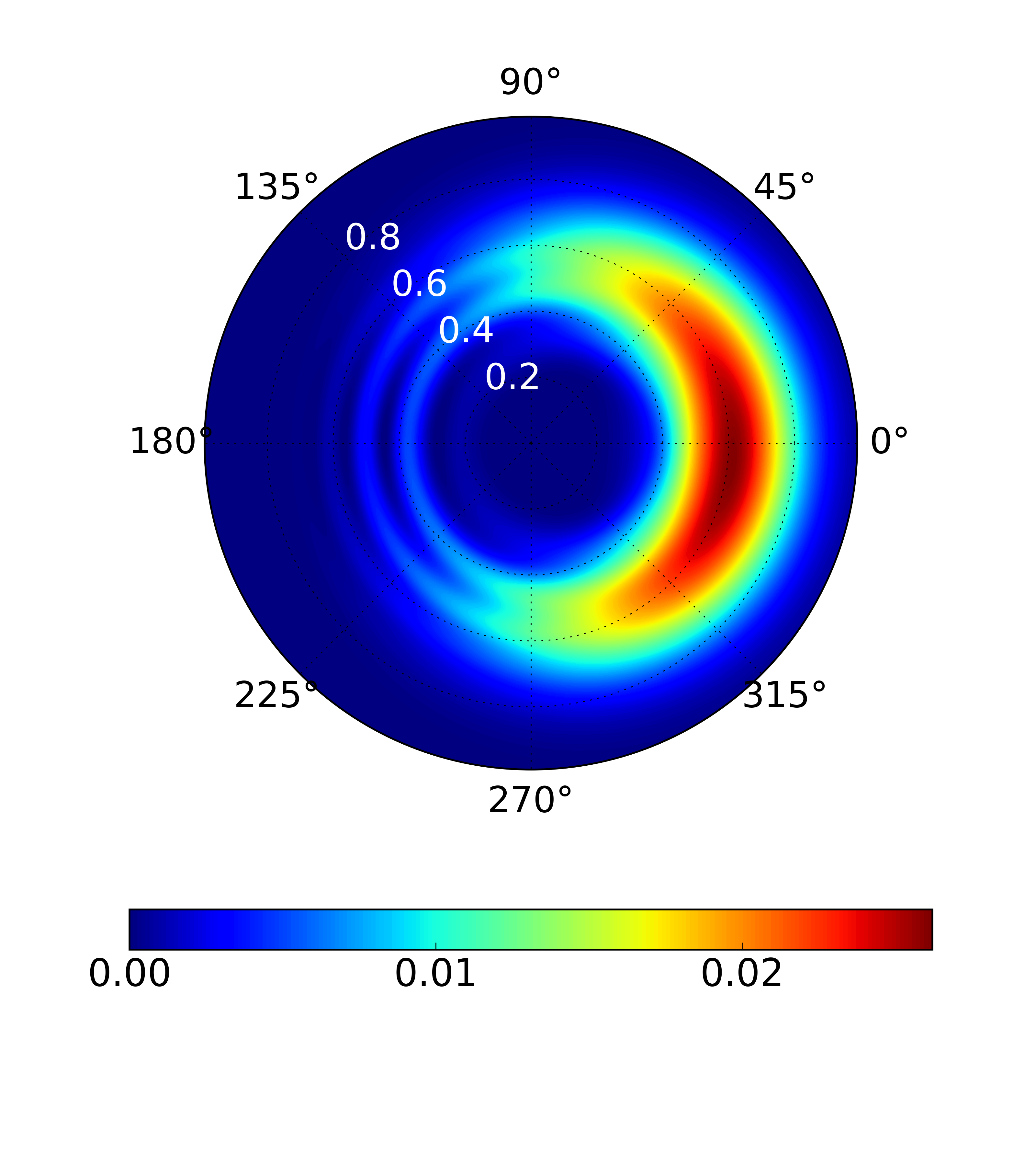}
&
        \includegraphics[width=0.5\columnwidth,trim={1cm 2cm 1cm 1cm}]{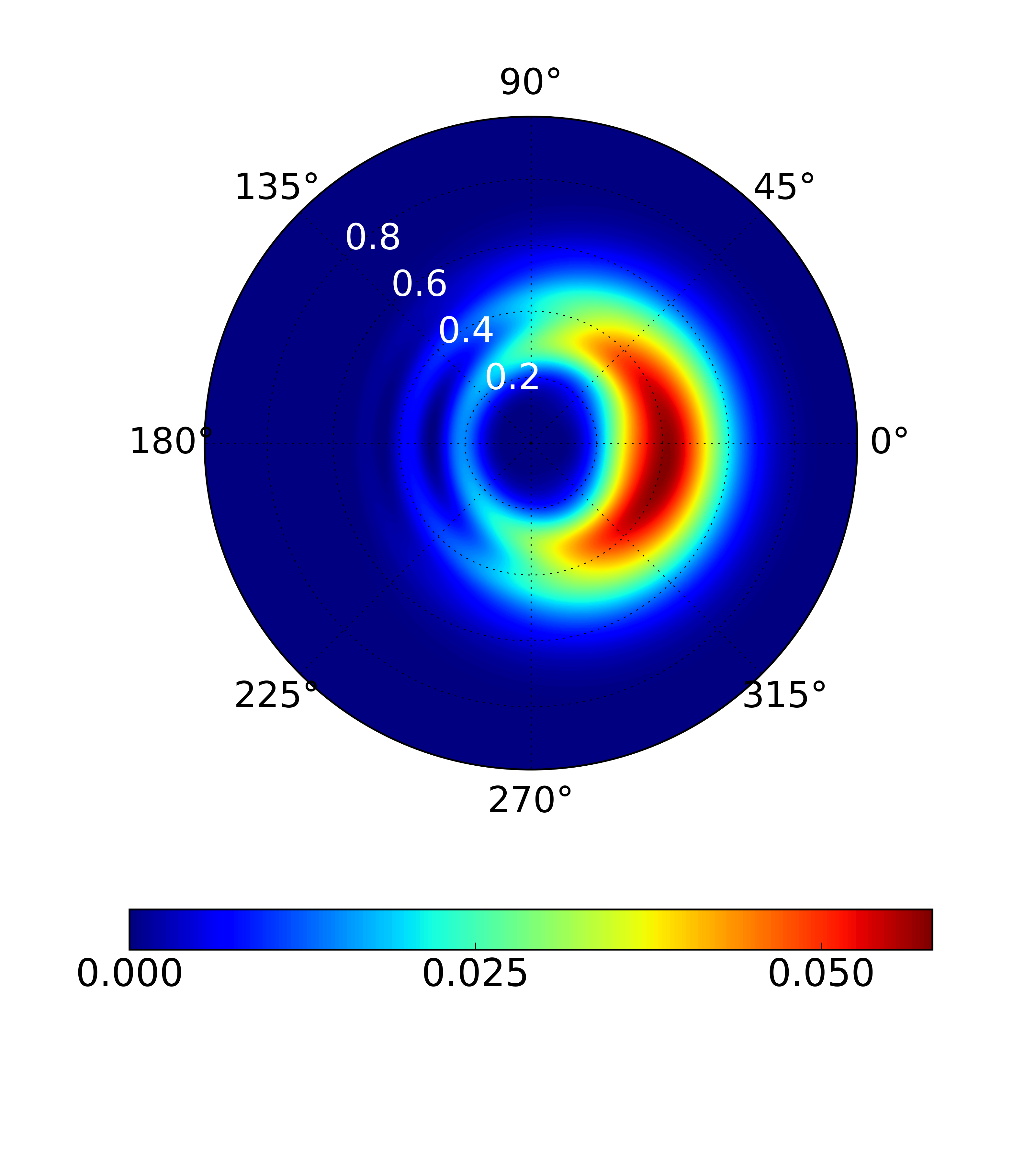}
\end{array}
$
\caption{Photoelectron momentum distributions in the polarization plane for (a) corotating and (b) counter-rotating electrons, arising from F\(^-\) in a two-cycle, 800-nm, \(3\times10^{13}\) W/cm\(^2\), circularly polarized laser pulse.}
\label{ppmplot}
\end{figure}

In circularly polarized fields, the detachment process is further complicated by the possible ejection of electrons from \(2p\) orbitals that are either corotating or counter-rotating with respect to the rotational sense of the laser field. We therefore investigate the offset angles for co- and counter-rotating electrons. Fig.\;\ref{ppmplot} shows their respective momentum distributions in the polarization plane. We note the expected dominance of counter-rotating electrons under these conditions, as well as the clear energy separation between the slow counter-rotating electrons and the faster corotating electrons. Both these tendencies have been seen in numerous calculations \cite{barth2011,kaushal2015,armstrong2019}. However, the relative angular offset within an attoclock scheme is less-thoroughly explored. In Refs.\;\cite{kaushal2015,kaushal2018b,liu2018}, it was found that under attoclock conditions co- and counter-rotating electrons in neutral atoms are separated not only in energy, but also in ejection angle. Their relative angular offset, typically of a few degrees, was attributed to the influence of the Coulomb potential, which deflects the slower counter-rotating electrons to a larger degree than the faster corotating ones. However, in a short-range potential, it was predicted that both co- and counter-rotating electrons ought to be ejected at zero degrees.

In Fig.\;\ref{ppmplot}, we demonstrate that significant angular offsets between co- and counter-rotating electrons can occur even in a short-range potential. We find that not only are both attoclock offsets non-zero, but there exists an angular separation between the co- and counter-rotating emissions. In the present case, at a peak intensity of \(3\times10^{13}\) W/cm\(^2\), corotating electrons display an offset of around --6 degrees, while counter-rotating electrons are offset by --9 degrees. We find that this relative offset between co- and counter-rotating electrons persists at other peak intensities. Furthermore, the offset angles are sensitive to the atomic structure description, with calculations using a Hartree-Fock model (not shown) again tending to give more negative angles in both cases. It is once more unclear how this effect would be manifest in a neutral system, since the differing momenta of co- and counter-rotating electrons will result in differing sensitivity to Coulomb deflections \cite{kaushal2015}. It may be that such deflections dominate in neutral systems, leading to the observation of larger positive offsets for counter-rotating electrons therein \cite{kaushal2015}.

\begin{figure}[t]
$
\begin{array}{cc}
(a) & (b)
\\
    \includegraphics[width=0.5\columnwidth,trim={1cm 2cm 1cm 1cm}]{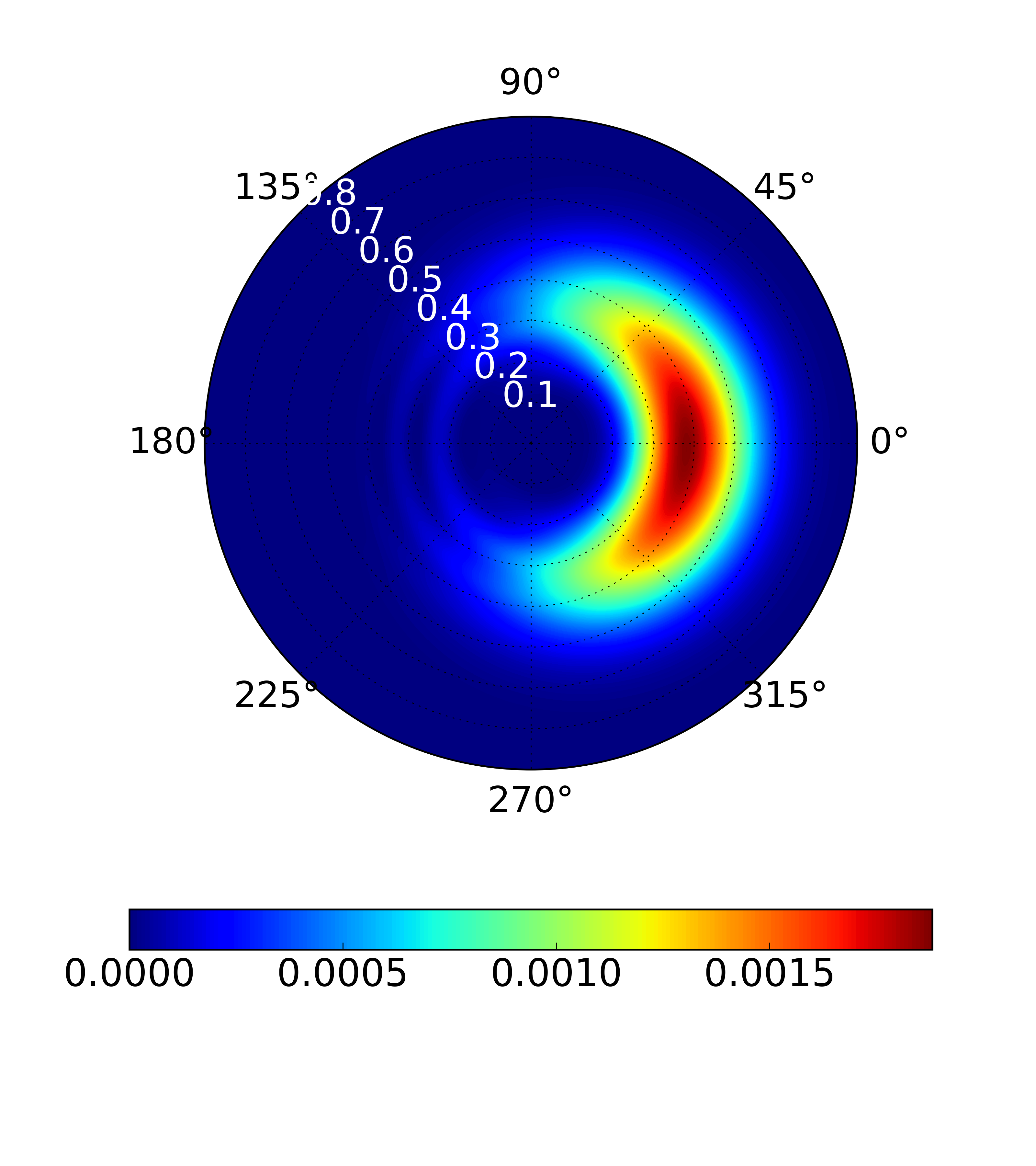}
    &
    \includegraphics[width=0.5\columnwidth,trim={1cm 2cm 1cm 1cm}]{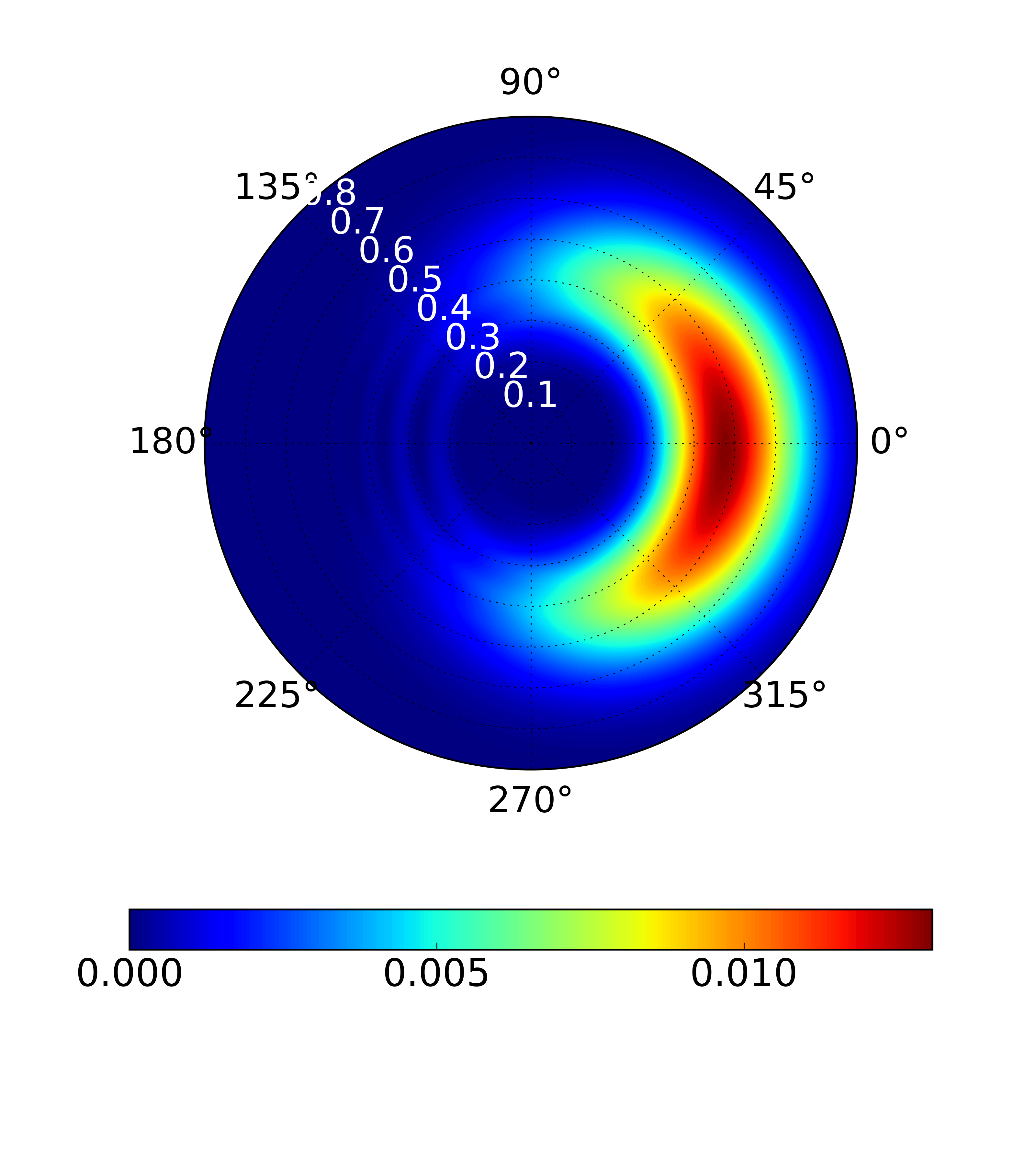}
\end{array}
$
\caption{Photoelectron momentum distributions in the polarization plane for F\(^-\), exposed to a two-cycle, 1500-nm, (a) \(5\times10^{12}\) W/cm\(^2\) and (b)  \(1\times10^{13}\) W/cm\(^2\) circularly polarized laser pulse.}
\label{fig:1500-1013}
\end{figure}

Finally, we seek to compare our calculations with those of Ref.\;\cite{douguet2019}, which treat a single active electron, and account for the induced dipole moment of the atomic residue through a time-dependent core polarization potential. We calculate the response of F\(^-\) to  1500-nm circularly polarized pulses of intensities \(5\times10^{12}\) W/cm\(^2\) and \(1\times10^{13}\) W/cm\(^2\). At this wavelength, at least five photons are required for electron detachment. The detachment probability therefore depends strongly on the peak intensity, and varies from 0.0009 at \(5\times10^{12}\) W/cm\(^2\) to 0.006 at  \(1\times10^{13}\) W/cm\(^2\). Fig.\;\ref{fig:1500-1013} shows the photoelectron momentum distribution in the laser polarization plane. Here, we find a common offset angle of around --2 degrees at both intensities. The SAE calculations for F\(^-\) under these conditions suggest negative offset angles between 0 degrees and --0.9 degrees  \cite{douguet2019}. Since both calculations naturally account for depletion effects, the discrepancy in offset angles must be attributable to the differing accounts of the short-range potential and core relaxation effects, where the latter, in particular, are absent in the SAE treatment.


In conclusion, we have investigated attosecond angular streaking of F$^-$ using the RMT method. Over a range of laser intensities and wavelengths, our {\em ab initio} calculations reveal strictly negative offset angles in the polarization-plane momentum distributions. 
We find that the calculated offset angles are sensitive to the quality of the atomic structure description of the core electrons.  This indicates that electron correlation can influence the offset angle in a multielectron system, despite having insignificant influence on a system such as helium. Further investigation of the detachment of co- and counter-rotating electrons reveals that both display negative offsets, and that a relative offset between the two is discernible. The slower counter-rotating electron displays a larger negative offset in all cases investigated. In comparison to SAE calculations, we find that our predicted offset angles differ by 1--2 degrees, and are never positive. This disparity is likely due to the differing treatments of the short-range potential, and associated core adjustments that take place during electron detachment. These effects are particularly visible due to the absence of a long-range Coulomb potential, and it is not clear to what extent they will be manifest in, for instance, noble-gas targets. Nonetheless, our results emphasize that core dynamics beyond the scope of SAE calculations can be present, and should be taken into account.

 The data presented in this article may be accessed at Ref.\;\cite{pure}. The RMT code is part of the UK-AMOR suite, and can be obtained for free at Ref.\;\cite{repo}. This work benefited from computational support by CoSeC, the Computational Science Centre for Research Communities, through CCPQ. The authors acknowledge funding from the EPSRC under grants EP/P022146/1, EP/P013953/1 and EP/R029342/1. This work relied on the ARCHER UK National Supercomputing Service (\url{www.archer.ac.uk}), for which access was obtained via the UK-AMOR consortium funded by EPSRC.

\appendix

\bibliography{mainbib}

\begin{thebibliography}{39}%
\makeatletter
\providecommand \@ifxundefined [1]{%
 \@ifx{#1\undefined}
}%
\providecommand \@ifnum [1]{%
 \ifnum #1\expandafter \@firstoftwo
 \else \expandafter \@secondoftwo
 \fi
}%
\providecommand \@ifx [1]{%
 \ifx #1\expandafter \@firstoftwo
 \else \expandafter \@secondoftwo
 \fi
}%
\providecommand \natexlab [1]{#1}%
\providecommand \enquote  [1]{``#1''}%
\providecommand \bibnamefont  [1]{#1}%
\providecommand \bibfnamefont [1]{#1}%
\providecommand \citenamefont [1]{#1}%
\providecommand \href@noop [0]{\@secondoftwo}%
\providecommand \href [0]{\begingroup \@sanitize@url \@href}%
\providecommand \@href[1]{\@@startlink{#1}\@@href}%
\providecommand \@@href[1]{\endgroup#1\@@endlink}%
\providecommand \@sanitize@url [0]{\catcode `\\12\catcode `\$12\catcode
  `\&12\catcode `\#12\catcode `\^12\catcode `\_12\catcode `\%12\relax}%
\providecommand \@@startlink[1]{}%
\providecommand \@@endlink[0]{}%
\providecommand \url  [0]{\begingroup\@sanitize@url \@url }%
\providecommand \@url [1]{\endgroup\@href {#1}{\urlprefix }}%
\providecommand \urlprefix  [0]{URL }%
\providecommand \Eprint [0]{\href }%
\providecommand \doibase [0]{http://dx.doi.org/}%
\providecommand \selectlanguage [0]{\@gobble}%
\providecommand \bibinfo  [0]{\@secondoftwo}%
\providecommand \bibfield  [0]{\@secondoftwo}%
\providecommand \translation [1]{[#1]}%
\providecommand \BibitemOpen [0]{}%
\providecommand \bibitemStop [0]{}%
\providecommand \bibitemNoStop [0]{.\EOS\space}%
\providecommand \EOS [0]{\spacefactor3000\relax}%
\providecommand \BibitemShut  [1]{\csname bibitem#1\endcsname}%
\let\auto@bib@innerbib\@empty
\bibitem [{\citenamefont {Eckle}\ \emph
  {et~al.}(2008{\natexlab{a}})\citenamefont {Eckle}, \citenamefont {Smolarski},
  \citenamefont {Schlup}, \citenamefont {Biegert}, \citenamefont {Staudte},
  \citenamefont {Sch{\"o}ffler}, \citenamefont {Muller}, \citenamefont
  {D{\"o}rner},\ and\ \citenamefont {Keller}}]{eckle2008}%
  \BibitemOpen
  \bibfield  {author} {\bibinfo {author} {\bibfnamefont {P.}~\bibnamefont
  {Eckle}}, \bibinfo {author} {\bibfnamefont {M.}~\bibnamefont {Smolarski}},
  \bibinfo {author} {\bibfnamefont {P.}~\bibnamefont {Schlup}}, \bibinfo
  {author} {\bibfnamefont {J.}~\bibnamefont {Biegert}}, \bibinfo {author}
  {\bibfnamefont {A.}~\bibnamefont {Staudte}}, \bibinfo {author} {\bibfnamefont
  {M.}~\bibnamefont {Sch{\"o}ffler}}, \bibinfo {author} {\bibfnamefont {H.~G.}\
  \bibnamefont {Muller}}, \bibinfo {author} {\bibfnamefont {R.}~\bibnamefont
  {D{\"o}rner}}, \ and\ \bibinfo {author} {\bibfnamefont {U.}~\bibnamefont
  {Keller}},\ }\href {https://doi.org/10.1038/nphys982} {\bibfield  {journal}
  {\bibinfo  {journal} {Nature Physics}\ }\textbf {\bibinfo {volume} {4}},\
  \bibinfo {pages} {565} (\bibinfo {year} {2008}{\natexlab{a}})}\BibitemShut
  {NoStop}%
\bibitem [{\citenamefont {Eckle}\ \emph
  {et~al.}(2008{\natexlab{b}})\citenamefont {Eckle}, \citenamefont {Pfeiffer},
  \citenamefont {Cirelli}, \citenamefont {Staudte}, \citenamefont {D{\"o}rner},
  \citenamefont {Muller}, \citenamefont {B{\"u}ttiker},\ and\ \citenamefont
  {Keller}}]{eckle2008sc}%
  \BibitemOpen
  \bibfield  {author} {\bibinfo {author} {\bibfnamefont {P.}~\bibnamefont
  {Eckle}}, \bibinfo {author} {\bibfnamefont {A.~N.}\ \bibnamefont {Pfeiffer}},
  \bibinfo {author} {\bibfnamefont {C.}~\bibnamefont {Cirelli}}, \bibinfo
  {author} {\bibfnamefont {A.}~\bibnamefont {Staudte}}, \bibinfo {author}
  {\bibfnamefont {R.}~\bibnamefont {D{\"o}rner}}, \bibinfo {author}
  {\bibfnamefont {H.~G.}\ \bibnamefont {Muller}}, \bibinfo {author}
  {\bibfnamefont {M.}~\bibnamefont {B{\"u}ttiker}}, \ and\ \bibinfo {author}
  {\bibfnamefont {U.}~\bibnamefont {Keller}},\ }\href {\doibase
  10.1126/science.1163439} {\bibfield  {journal} {\bibinfo  {journal}
  {Science}\ }\textbf {\bibinfo {volume} {322}},\ \bibinfo {pages} {1525}
  (\bibinfo {year} {2008}{\natexlab{b}})}\BibitemShut {NoStop}%
\bibitem [{\citenamefont {Pfeiffer}\ \emph {et~al.}(2011)\citenamefont
  {Pfeiffer}, \citenamefont {Cirelli}, \citenamefont {Smolarski}, \citenamefont
  {Dimitrovski}, \citenamefont {Abu-samha}, \citenamefont {Madsen},\ and\
  \citenamefont {Keller}}]{pfeiffer2012}%
  \BibitemOpen
  \bibfield  {author} {\bibinfo {author} {\bibfnamefont {A.~N.}\ \bibnamefont
  {Pfeiffer}}, \bibinfo {author} {\bibfnamefont {C.}~\bibnamefont {Cirelli}},
  \bibinfo {author} {\bibfnamefont {M.}~\bibnamefont {Smolarski}}, \bibinfo
  {author} {\bibfnamefont {D.}~\bibnamefont {Dimitrovski}}, \bibinfo {author}
  {\bibfnamefont {M.}~\bibnamefont {Abu-samha}}, \bibinfo {author}
  {\bibfnamefont {L.~B.}\ \bibnamefont {Madsen}}, \ and\ \bibinfo {author}
  {\bibfnamefont {U.}~\bibnamefont {Keller}},\ }\href
  {https://doi.org/10.1038/nphys2125} {\bibfield  {journal} {\bibinfo
  {journal} {Nature Physics}\ }\textbf {\bibinfo {volume} {8}},\ \bibinfo
  {pages} {76} (\bibinfo {year} {2011})}\BibitemShut {NoStop}%
\bibitem [{\citenamefont {Landsman}\ \emph {et~al.}(2014)\citenamefont
  {Landsman}, \citenamefont {Weger}, \citenamefont {Maurer}, \citenamefont
  {Boge}, \citenamefont {Ludwig}, \citenamefont {Heuser}, \citenamefont
  {Cirelli}, \citenamefont {Gallmann},\ and\ \citenamefont
  {Keller}}]{landsman2014}%
  \BibitemOpen
  \bibfield  {author} {\bibinfo {author} {\bibfnamefont {A.~S.}\ \bibnamefont
  {Landsman}}, \bibinfo {author} {\bibfnamefont {M.}~\bibnamefont {Weger}},
  \bibinfo {author} {\bibfnamefont {J.}~\bibnamefont {Maurer}}, \bibinfo
  {author} {\bibfnamefont {R.}~\bibnamefont {Boge}}, \bibinfo {author}
  {\bibfnamefont {A.}~\bibnamefont {Ludwig}}, \bibinfo {author} {\bibfnamefont
  {S.}~\bibnamefont {Heuser}}, \bibinfo {author} {\bibfnamefont
  {C.}~\bibnamefont {Cirelli}}, \bibinfo {author} {\bibfnamefont
  {L.}~\bibnamefont {Gallmann}}, \ and\ \bibinfo {author} {\bibfnamefont
  {U.}~\bibnamefont {Keller}},\ }\href {\doibase 10.1364/OPTICA.1.000343}
  {\bibfield  {journal} {\bibinfo  {journal} {Optica}\ }\textbf {\bibinfo
  {volume} {1}},\ \bibinfo {pages} {343} (\bibinfo {year} {2014})}\BibitemShut
  {NoStop}%
\bibitem [{\citenamefont {Camus}\ \emph {et~al.}(2017)\citenamefont {Camus},
  \citenamefont {Yakaboylu}, \citenamefont {Fechner}, \citenamefont {Klaiber},
  \citenamefont {Laux}, \citenamefont {Mi}, \citenamefont {Hatsagortsyan},
  \citenamefont {Pfeifer}, \citenamefont {Keitel},\ and\ \citenamefont
  {Moshammer}}]{camus2017}%
  \BibitemOpen
  \bibfield  {author} {\bibinfo {author} {\bibfnamefont {N.}~\bibnamefont
  {Camus}}, \bibinfo {author} {\bibfnamefont {E.}~\bibnamefont {Yakaboylu}},
  \bibinfo {author} {\bibfnamefont {L.}~\bibnamefont {Fechner}}, \bibinfo
  {author} {\bibfnamefont {M.}~\bibnamefont {Klaiber}}, \bibinfo {author}
  {\bibfnamefont {M.}~\bibnamefont {Laux}}, \bibinfo {author} {\bibfnamefont
  {Y.}~\bibnamefont {Mi}}, \bibinfo {author} {\bibfnamefont {K.~Z.}\
  \bibnamefont {Hatsagortsyan}}, \bibinfo {author} {\bibfnamefont
  {T.}~\bibnamefont {Pfeifer}}, \bibinfo {author} {\bibfnamefont {C.~H.}\
  \bibnamefont {Keitel}}, \ and\ \bibinfo {author} {\bibfnamefont
  {R.}~\bibnamefont {Moshammer}},\ }\href {\doibase
  10.1103/PhysRevLett.119.023201} {\bibfield  {journal} {\bibinfo  {journal}
  {Phys. Rev. Lett.}\ }\textbf {\bibinfo {volume} {119}},\ \bibinfo {pages}
  {023201} (\bibinfo {year} {2017})}\BibitemShut {NoStop}%
\bibitem [{\citenamefont {Fortun}\ \emph {et~al.}(2016)\citenamefont {Fortun},
  \citenamefont {Cabrera-Guti\'errez}, \citenamefont {Condon}, \citenamefont
  {Michon}, \citenamefont {Billy},\ and\ \citenamefont
  {Gu\'ery-Odelin}}]{fortun2016}%
  \BibitemOpen
  \bibfield  {author} {\bibinfo {author} {\bibfnamefont {A.}~\bibnamefont
  {Fortun}}, \bibinfo {author} {\bibfnamefont {C.}~\bibnamefont
  {Cabrera-Guti\'errez}}, \bibinfo {author} {\bibfnamefont {G.}~\bibnamefont
  {Condon}}, \bibinfo {author} {\bibfnamefont {E.}~\bibnamefont {Michon}},
  \bibinfo {author} {\bibfnamefont {J.}~\bibnamefont {Billy}}, \ and\ \bibinfo
  {author} {\bibfnamefont {D.}~\bibnamefont {Gu\'ery-Odelin}},\ }\href
  {\doibase 10.1103/PhysRevLett.117.010401} {\bibfield  {journal} {\bibinfo
  {journal} {Phys. Rev. Lett.}\ }\textbf {\bibinfo {volume} {117}},\ \bibinfo
  {pages} {010401} (\bibinfo {year} {2016})}\BibitemShut {NoStop}%
\bibitem [{\citenamefont {Ni}\ \emph {et~al.}(2016)\citenamefont {Ni},
  \citenamefont {Saalmann},\ and\ \citenamefont {Rost}}]{ni2016}%
  \BibitemOpen
  \bibfield  {author} {\bibinfo {author} {\bibfnamefont {H.}~\bibnamefont
  {Ni}}, \bibinfo {author} {\bibfnamefont {U.}~\bibnamefont {Saalmann}}, \ and\
  \bibinfo {author} {\bibfnamefont {J.-M.}\ \bibnamefont {Rost}},\ }\href
  {\doibase 10.1103/PhysRevLett.117.023002} {\bibfield  {journal} {\bibinfo
  {journal} {Phys. Rev. Lett.}\ }\textbf {\bibinfo {volume} {117}},\ \bibinfo
  {pages} {023002} (\bibinfo {year} {2016})}\BibitemShut {NoStop}%
\bibitem [{\citenamefont {Torlina}\ \emph {et~al.}(2015)\citenamefont
  {Torlina}, \citenamefont {Morales}, \citenamefont {Kaushal}, \citenamefont
  {Ivanov}, \citenamefont {Kheifets}, \citenamefont {Zielinski}, \citenamefont
  {Scrinzi}, \citenamefont {Muller}, \citenamefont {Sukiasyan}, \citenamefont
  {Ivanov},\ and\ \citenamefont {Smirnova}}]{torlina2015}%
  \BibitemOpen
  \bibfield  {author} {\bibinfo {author} {\bibfnamefont {L.}~\bibnamefont
  {Torlina}}, \bibinfo {author} {\bibfnamefont {F.}~\bibnamefont {Morales}},
  \bibinfo {author} {\bibfnamefont {J.}~\bibnamefont {Kaushal}}, \bibinfo
  {author} {\bibfnamefont {I.}~\bibnamefont {Ivanov}}, \bibinfo {author}
  {\bibfnamefont {A.}~\bibnamefont {Kheifets}}, \bibinfo {author}
  {\bibfnamefont {A.}~\bibnamefont {Zielinski}}, \bibinfo {author}
  {\bibfnamefont {A.}~\bibnamefont {Scrinzi}}, \bibinfo {author} {\bibfnamefont
  {H.~G.}\ \bibnamefont {Muller}}, \bibinfo {author} {\bibfnamefont
  {S.}~\bibnamefont {Sukiasyan}}, \bibinfo {author} {\bibfnamefont
  {M.}~\bibnamefont {Ivanov}}, \ and\ \bibinfo {author} {\bibfnamefont
  {O.}~\bibnamefont {Smirnova}},\ }\href {https://doi.org/10.1038/nphys3340}
  {\bibfield  {journal} {\bibinfo  {journal} {Nature Physics}\ }\textbf
  {\bibinfo {volume} {11}},\ \bibinfo {pages} {503} (\bibinfo {year}
  {2015})}\BibitemShut {NoStop}%
\bibitem [{\citenamefont {Bray}\ \emph {et~al.}(2018)\citenamefont {Bray},
  \citenamefont {Eckart},\ and\ \citenamefont {Kheifets}}]{bray2018}%
  \BibitemOpen
  \bibfield  {author} {\bibinfo {author} {\bibfnamefont {A.~W.}\ \bibnamefont
  {Bray}}, \bibinfo {author} {\bibfnamefont {S.}~\bibnamefont {Eckart}}, \ and\
  \bibinfo {author} {\bibfnamefont {A.~S.}\ \bibnamefont {Kheifets}},\ }\href
  {\doibase 10.1103/PhysRevLett.121.123201} {\bibfield  {journal} {\bibinfo
  {journal} {Phys. Rev. Lett.}\ }\textbf {\bibinfo {volume} {121}},\ \bibinfo
  {pages} {123201} (\bibinfo {year} {2018})}\BibitemShut {NoStop}%
\bibitem [{\citenamefont {Sainadh}\ \emph {et~al.}(2019)\citenamefont
  {Sainadh}, \citenamefont {Xu}, \citenamefont {Wang}, \citenamefont
  {Atia-Tul-Noor}, \citenamefont {Wallace}, \citenamefont {Douguet},
  \citenamefont {Bray}, \citenamefont {Ivanov}, \citenamefont {Bartschat},
  \citenamefont {Kheifets}, \citenamefont {Sang},\ and\ \citenamefont
  {Litvinyuk}}]{sainadh2019}%
  \BibitemOpen
  \bibfield  {author} {\bibinfo {author} {\bibfnamefont {U.~S.}\ \bibnamefont
  {Sainadh}}, \bibinfo {author} {\bibfnamefont {H.}~\bibnamefont {Xu}},
  \bibinfo {author} {\bibfnamefont {X.}~\bibnamefont {Wang}}, \bibinfo {author}
  {\bibfnamefont {A.}~\bibnamefont {Atia-Tul-Noor}}, \bibinfo {author}
  {\bibfnamefont {W.~C.}\ \bibnamefont {Wallace}}, \bibinfo {author}
  {\bibfnamefont {N.}~\bibnamefont {Douguet}}, \bibinfo {author} {\bibfnamefont
  {A.}~\bibnamefont {Bray}}, \bibinfo {author} {\bibfnamefont {I.}~\bibnamefont
  {Ivanov}}, \bibinfo {author} {\bibfnamefont {K.}~\bibnamefont {Bartschat}},
  \bibinfo {author} {\bibfnamefont {A.}~\bibnamefont {Kheifets}}, \bibinfo
  {author} {\bibfnamefont {R.~T.}\ \bibnamefont {Sang}}, \ and\ \bibinfo
  {author} {\bibfnamefont {I.~V.}\ \bibnamefont {Litvinyuk}},\ }\href {\doibase
  10.1038/s41586-019-1028-3} {\bibfield  {journal} {\bibinfo  {journal}
  {Nature}\ }\textbf {\bibinfo {volume} {568}},\ \bibinfo {pages} {75}
  (\bibinfo {year} {2019})}\BibitemShut {NoStop}%
\bibitem [{\citenamefont {Hofmann}\ \emph {et~al.}(2019)\citenamefont
  {Hofmann}, \citenamefont {Landsman},\ and\ \citenamefont
  {Keller}}]{hofmann2019}%
  \BibitemOpen
  \bibfield  {author} {\bibinfo {author} {\bibfnamefont {C.}~\bibnamefont
  {Hofmann}}, \bibinfo {author} {\bibfnamefont {A.~S.}\ \bibnamefont
  {Landsman}}, \ and\ \bibinfo {author} {\bibfnamefont {U.}~\bibnamefont
  {Keller}},\ }\href {\doibase 10.1080/09500340.2019.1596325} {\bibfield
  {journal} {\bibinfo  {journal} {Journal of Modern Optics}\ }\textbf {\bibinfo
  {volume} {66}},\ \bibinfo {pages} {1052} (\bibinfo {year}
  {2019})}\BibitemShut {NoStop}%
\bibitem [{\citenamefont {Teeny}\ \emph {et~al.}(2016)\citenamefont {Teeny},
  \citenamefont {Yakaboylu}, \citenamefont {Bauke},\ and\ \citenamefont
  {Keitel}}]{teeny2016}%
  \BibitemOpen
  \bibfield  {author} {\bibinfo {author} {\bibfnamefont {N.}~\bibnamefont
  {Teeny}}, \bibinfo {author} {\bibfnamefont {E.}~\bibnamefont {Yakaboylu}},
  \bibinfo {author} {\bibfnamefont {H.}~\bibnamefont {Bauke}}, \ and\ \bibinfo
  {author} {\bibfnamefont {C.~H.}\ \bibnamefont {Keitel}},\ }\href {\doibase
  10.1103/PhysRevLett.116.063003} {\bibfield  {journal} {\bibinfo  {journal}
  {Phys. Rev. Lett.}\ }\textbf {\bibinfo {volume} {116}},\ \bibinfo {pages}
  {063003} (\bibinfo {year} {2016})}\BibitemShut {NoStop}%
\bibitem [{\citenamefont {Boge}\ \emph {et~al.}(2013)\citenamefont {Boge},
  \citenamefont {Cirelli}, \citenamefont {Landsman}, \citenamefont {Heuser},
  \citenamefont {Ludwig}, \citenamefont {Maurer}, \citenamefont {Weger},
  \citenamefont {Gallmann},\ and\ \citenamefont {Keller}}]{boge2013}%
  \BibitemOpen
  \bibfield  {author} {\bibinfo {author} {\bibfnamefont {R.}~\bibnamefont
  {Boge}}, \bibinfo {author} {\bibfnamefont {C.}~\bibnamefont {Cirelli}},
  \bibinfo {author} {\bibfnamefont {A.~S.}\ \bibnamefont {Landsman}}, \bibinfo
  {author} {\bibfnamefont {S.}~\bibnamefont {Heuser}}, \bibinfo {author}
  {\bibfnamefont {A.}~\bibnamefont {Ludwig}}, \bibinfo {author} {\bibfnamefont
  {J.}~\bibnamefont {Maurer}}, \bibinfo {author} {\bibfnamefont
  {M.}~\bibnamefont {Weger}}, \bibinfo {author} {\bibfnamefont
  {L.}~\bibnamefont {Gallmann}}, \ and\ \bibinfo {author} {\bibfnamefont
  {U.}~\bibnamefont {Keller}},\ }\href {\doibase
  10.1103/PhysRevLett.111.103003} {\bibfield  {journal} {\bibinfo  {journal}
  {Phys. Rev. Lett.}\ }\textbf {\bibinfo {volume} {111}},\ \bibinfo {pages}
  {103003} (\bibinfo {year} {2013})}\BibitemShut {NoStop}%
\bibitem [{\citenamefont {Ivanov}\ and\ \citenamefont
  {Kheifets}(2014)}]{ivanov2014}%
  \BibitemOpen
  \bibfield  {author} {\bibinfo {author} {\bibfnamefont {I.~A.}\ \bibnamefont
  {Ivanov}}\ and\ \bibinfo {author} {\bibfnamefont {A.~S.}\ \bibnamefont
  {Kheifets}},\ }\href {\doibase 10.1103/PhysRevA.89.021402} {\bibfield
  {journal} {\bibinfo  {journal} {Phys. Rev. A}\ }\textbf {\bibinfo {volume}
  {89}},\ \bibinfo {pages} {021402} (\bibinfo {year} {2014})}\BibitemShut
  {NoStop}%
\bibitem [{\citenamefont {Hofmann}\ \emph {et~al.}(2014)\citenamefont
  {Hofmann}, \citenamefont {Landsman}, \citenamefont {Zielinski}, \citenamefont
  {Cirelli}, \citenamefont {Zimmermann}, \citenamefont {Scrinzi},\ and\
  \citenamefont {Keller}}]{hofmann2014}%
  \BibitemOpen
  \bibfield  {author} {\bibinfo {author} {\bibfnamefont {C.}~\bibnamefont
  {Hofmann}}, \bibinfo {author} {\bibfnamefont {A.~S.}\ \bibnamefont
  {Landsman}}, \bibinfo {author} {\bibfnamefont {A.}~\bibnamefont {Zielinski}},
  \bibinfo {author} {\bibfnamefont {C.}~\bibnamefont {Cirelli}}, \bibinfo
  {author} {\bibfnamefont {T.}~\bibnamefont {Zimmermann}}, \bibinfo {author}
  {\bibfnamefont {A.}~\bibnamefont {Scrinzi}}, \ and\ \bibinfo {author}
  {\bibfnamefont {U.}~\bibnamefont {Keller}},\ }\href {\doibase
  10.1103/PhysRevA.90.043406} {\bibfield  {journal} {\bibinfo  {journal} {Phys.
  Rev. A}\ }\textbf {\bibinfo {volume} {90}},\ \bibinfo {pages} {043406}
  (\bibinfo {year} {2014})}\BibitemShut {NoStop}%
\bibitem [{\citenamefont {Klaiber}\ \emph {et~al.}(2015)\citenamefont
  {Klaiber}, \citenamefont {Hatsagortsyan},\ and\ \citenamefont
  {Keitel}}]{klaiber2015}%
  \BibitemOpen
  \bibfield  {author} {\bibinfo {author} {\bibfnamefont {M.}~\bibnamefont
  {Klaiber}}, \bibinfo {author} {\bibfnamefont {K.~Z.}\ \bibnamefont
  {Hatsagortsyan}}, \ and\ \bibinfo {author} {\bibfnamefont {C.~H.}\
  \bibnamefont {Keitel}},\ }\href {\doibase 10.1103/PhysRevLett.114.083001}
  {\bibfield  {journal} {\bibinfo  {journal} {Phys. Rev. Lett.}\ }\textbf
  {\bibinfo {volume} {114}},\ \bibinfo {pages} {083001} (\bibinfo {year}
  {2015})}\BibitemShut {NoStop}%
\bibitem [{\citenamefont {Han}\ \emph {et~al.}(2019)\citenamefont {Han},
  \citenamefont {Ge}, \citenamefont {Fang}, \citenamefont {Yu}, \citenamefont
  {Guo}, \citenamefont {Ma}, \citenamefont {Deng}, \citenamefont {Gong},\ and\
  \citenamefont {Liu}}]{han2019}%
  \BibitemOpen
  \bibfield  {author} {\bibinfo {author} {\bibfnamefont {M.}~\bibnamefont
  {Han}}, \bibinfo {author} {\bibfnamefont {P.}~\bibnamefont {Ge}}, \bibinfo
  {author} {\bibfnamefont {Y.}~\bibnamefont {Fang}}, \bibinfo {author}
  {\bibfnamefont {X.}~\bibnamefont {Yu}}, \bibinfo {author} {\bibfnamefont
  {Z.}~\bibnamefont {Guo}}, \bibinfo {author} {\bibfnamefont {X.}~\bibnamefont
  {Ma}}, \bibinfo {author} {\bibfnamefont {Y.}~\bibnamefont {Deng}}, \bibinfo
  {author} {\bibfnamefont {Q.}~\bibnamefont {Gong}}, \ and\ \bibinfo {author}
  {\bibfnamefont {Y.}~\bibnamefont {Liu}},\ }\href {\doibase
  10.1103/PhysRevLett.123.073201} {\bibfield  {journal} {\bibinfo  {journal}
  {Phys. Rev. Lett.}\ }\textbf {\bibinfo {volume} {123}},\ \bibinfo {pages}
  {073201} (\bibinfo {year} {2019})}\BibitemShut {NoStop}%
\bibitem [{\citenamefont {Eicke}\ \emph {et~al.}(2020)\citenamefont {Eicke},
  \citenamefont {Brennecke},\ and\ \citenamefont {Lein}}]{eicke2020}%
  \BibitemOpen
  \bibfield  {author} {\bibinfo {author} {\bibfnamefont {N.}~\bibnamefont
  {Eicke}}, \bibinfo {author} {\bibfnamefont {S.}~\bibnamefont {Brennecke}}, \
  and\ \bibinfo {author} {\bibfnamefont {M.}~\bibnamefont {Lein}},\ }\href
  {\doibase 10.1103/PhysRevLett.124.043202} {\bibfield  {journal} {\bibinfo
  {journal} {Phys. Rev. Lett.}\ }\textbf {\bibinfo {volume} {124}},\ \bibinfo
  {pages} {043202} (\bibinfo {year} {2020})}\BibitemShut {NoStop}%
\bibitem [{\citenamefont {Ni}\ \emph {et~al.}(2018)\citenamefont {Ni},
  \citenamefont {Saalmann},\ and\ \citenamefont {Rost}}]{ni2018}%
  \BibitemOpen
  \bibfield  {author} {\bibinfo {author} {\bibfnamefont {H.}~\bibnamefont
  {Ni}}, \bibinfo {author} {\bibfnamefont {U.}~\bibnamefont {Saalmann}}, \ and\
  \bibinfo {author} {\bibfnamefont {J.-M.}\ \bibnamefont {Rost}},\ }\href
  {\doibase 10.1103/PhysRevA.97.013426} {\bibfield  {journal} {\bibinfo
  {journal} {Phys. Rev. A}\ }\textbf {\bibinfo {volume} {97}},\ \bibinfo
  {pages} {013426} (\bibinfo {year} {2018})}\BibitemShut {NoStop}%
\bibitem [{\citenamefont {Douguet}\ and\ \citenamefont
  {Bartschat}(2018)}]{douguet2018}%
  \BibitemOpen
  \bibfield  {author} {\bibinfo {author} {\bibfnamefont {N.}~\bibnamefont
  {Douguet}}\ and\ \bibinfo {author} {\bibfnamefont {K.}~\bibnamefont
  {Bartschat}},\ }\href {\doibase 10.1103/PhysRevA.97.013402} {\bibfield
  {journal} {\bibinfo  {journal} {Phys. Rev. A}\ }\textbf {\bibinfo {volume}
  {97}},\ \bibinfo {pages} {013402} (\bibinfo {year} {2018})}\BibitemShut
  {NoStop}%
\bibitem [{\citenamefont {Douguet}\ and\ \citenamefont
  {Bartschat}(2019)}]{douguet2019}%
  \BibitemOpen
  \bibfield  {author} {\bibinfo {author} {\bibfnamefont {N.}~\bibnamefont
  {Douguet}}\ and\ \bibinfo {author} {\bibfnamefont {K.}~\bibnamefont
  {Bartschat}},\ }\href {\doibase 10.1103/PhysRevA.99.023417} {\bibfield
  {journal} {\bibinfo  {journal} {Phys. Rev. A}\ }\textbf {\bibinfo {volume}
  {99}},\ \bibinfo {pages} {023417} (\bibinfo {year} {2019})}\BibitemShut
  {NoStop}%
\bibitem [{\citenamefont {Emmanouilidou}\ \emph {et~al.}(2015)\citenamefont
  {Emmanouilidou}, \citenamefont {Chen}, \citenamefont {Hofmann}, \citenamefont
  {Keller},\ and\ \citenamefont {Landsman}}]{agapi2015}%
  \BibitemOpen
  \bibfield  {author} {\bibinfo {author} {\bibfnamefont {A.}~\bibnamefont
  {Emmanouilidou}}, \bibinfo {author} {\bibfnamefont {A.}~\bibnamefont {Chen}},
  \bibinfo {author} {\bibfnamefont {C.}~\bibnamefont {Hofmann}}, \bibinfo
  {author} {\bibfnamefont {U.}~\bibnamefont {Keller}}, \ and\ \bibinfo {author}
  {\bibfnamefont {A.~S.}\ \bibnamefont {Landsman}},\ }\href {\doibase
  10.1088/0953-4075/48/24/245602} {\bibfield  {journal} {\bibinfo  {journal}
  {Journal of Physics B: Atomic, Molecular and Optical Physics}\ }\textbf
  {\bibinfo {volume} {48}},\ \bibinfo {pages} {245602} (\bibinfo {year}
  {2015})}\BibitemShut {NoStop}%
\bibitem [{\citenamefont {Majety}\ and\ \citenamefont
  {Scrinzi}(2017)}]{majety2017}%
  \BibitemOpen
  \bibfield  {author} {\bibinfo {author} {\bibfnamefont {V.~P.}\ \bibnamefont
  {Majety}}\ and\ \bibinfo {author} {\bibfnamefont {A.}~\bibnamefont
  {Scrinzi}},\ }\href {\doibase 10.1080/09500340.2016.1271915} {\bibfield
  {journal} {\bibinfo  {journal} {Journal of Modern Optics}\ }\textbf {\bibinfo
  {volume} {64}},\ \bibinfo {pages} {1026} (\bibinfo {year}
  {2017})}\BibitemShut {NoStop}%
\bibitem [{\citenamefont {Nikolopoulos}\ \emph {et~al.}(2008)\citenamefont
  {Nikolopoulos}, \citenamefont {Parker},\ and\ \citenamefont
  {Taylor}}]{nikolopoulos2008}%
  \BibitemOpen
  \bibfield  {author} {\bibinfo {author} {\bibfnamefont {L.~A.~A.}\
  \bibnamefont {Nikolopoulos}}, \bibinfo {author} {\bibfnamefont {J.~S.}\
  \bibnamefont {Parker}}, \ and\ \bibinfo {author} {\bibfnamefont {K.~T.}\
  \bibnamefont {Taylor}},\ }\href {\doibase 10.1103/PhysRevA.78.063420}
  {\bibfield  {journal} {\bibinfo  {journal} {Phys. Rev. A}\ }\textbf {\bibinfo
  {volume} {78}},\ \bibinfo {pages} {063420} (\bibinfo {year}
  {2008})}\BibitemShut {NoStop}%
\bibitem [{\citenamefont {Moore}\ \emph {et~al.}(2011)\citenamefont {Moore},
  \citenamefont {Lysaght}, \citenamefont {Nikolopoulos}, \citenamefont
  {Parker}, \citenamefont {van~der Hart},\ and\ \citenamefont
  {Taylor}}]{moore2011}%
  \BibitemOpen
  \bibfield  {author} {\bibinfo {author} {\bibfnamefont {L.~R.}\ \bibnamefont
  {Moore}}, \bibinfo {author} {\bibfnamefont {M.~A.}\ \bibnamefont {Lysaght}},
  \bibinfo {author} {\bibfnamefont {L.~A.~A.}\ \bibnamefont {Nikolopoulos}},
  \bibinfo {author} {\bibfnamefont {J.~S.}\ \bibnamefont {Parker}}, \bibinfo
  {author} {\bibfnamefont {H.~W.}\ \bibnamefont {van~der Hart}}, \ and\
  \bibinfo {author} {\bibfnamefont {K.~T.}\ \bibnamefont {Taylor}},\ }\href
  {https://doi.org/10.1080/09500340.2011.559315} {\bibfield  {journal}
  {\bibinfo  {journal} {Journal of Modern Optics}\ }\textbf {\bibinfo {volume}
  {58}},\ \bibinfo {pages} {1132} (\bibinfo {year} {2011})}\BibitemShut
  {NoStop}%
\bibitem [{\citenamefont {Clarke}\ \emph {et~al.}(2018)\citenamefont {Clarke},
  \citenamefont {Armstrong}, \citenamefont {Brown},\ and\ \citenamefont
  {van~der Hart}}]{clarke2018}%
  \BibitemOpen
  \bibfield  {author} {\bibinfo {author} {\bibfnamefont {D.~D.~A.}\
  \bibnamefont {Clarke}}, \bibinfo {author} {\bibfnamefont {G.~S.~J.}\
  \bibnamefont {Armstrong}}, \bibinfo {author} {\bibfnamefont {A.~C.}\
  \bibnamefont {Brown}}, \ and\ \bibinfo {author} {\bibfnamefont {H.~W.}\
  \bibnamefont {van~der Hart}},\ }\href {\doibase 10.1103/PhysRevA.98.053442}
  {\bibfield  {journal} {\bibinfo  {journal} {Phys. Rev. A}\ }\textbf {\bibinfo
  {volume} {98}},\ \bibinfo {pages} {053442} (\bibinfo {year}
  {2018})}\BibitemShut {NoStop}%
\bibitem [{\citenamefont {Brown}\ \emph {et~al.}(2020)\citenamefont {Brown},
  \citenamefont {Armstrong}, \citenamefont {Benda}, \citenamefont {Clarke},
  \citenamefont {Wragg}, \citenamefont {Hamilton}, \citenamefont
  {Ma\v{s}\'{i}n}, \citenamefont {Gorfinkiel},\ and\ \citenamefont {van~der
  Hart}}]{rmtcpc}%
  \BibitemOpen
  \bibfield  {author} {\bibinfo {author} {\bibfnamefont {A.~C.}\ \bibnamefont
  {Brown}}, \bibinfo {author} {\bibfnamefont {G.~S.~J.}\ \bibnamefont
  {Armstrong}}, \bibinfo {author} {\bibfnamefont {J.}~\bibnamefont {Benda}},
  \bibinfo {author} {\bibfnamefont {D.~D.~A.}\ \bibnamefont {Clarke}}, \bibinfo
  {author} {\bibfnamefont {J.}~\bibnamefont {Wragg}}, \bibinfo {author}
  {\bibfnamefont {K.~R.}\ \bibnamefont {Hamilton}}, \bibinfo {author}
  {\bibfnamefont {Z.}~\bibnamefont {Ma\v{s}\'{i}n}}, \bibinfo {author}
  {\bibfnamefont {J.~D.}\ \bibnamefont {Gorfinkiel}}, \ and\ \bibinfo {author}
  {\bibfnamefont {H.~W.}\ \bibnamefont {van~der Hart}},\ }\href {\doibase
  https://doi.org/10.1016/j.cpc.2019.107062} {\bibfield  {journal} {\bibinfo
  {journal} {Computer Physics Communications}\ }\textbf {\bibinfo {volume}
  {250}},\ \bibinfo {pages} {107062} (\bibinfo {year} {2020})}\BibitemShut
  {NoStop}%
\bibitem [{\citenamefont {Hassouneh}\ \emph {et~al.}(2015)\citenamefont
  {Hassouneh}, \citenamefont {Law}, \citenamefont {Shearer}, \citenamefont
  {Brown},\ and\ \citenamefont {van~der Hart}}]{hassouneh2015}%
  \BibitemOpen
  \bibfield  {author} {\bibinfo {author} {\bibfnamefont {O.}~\bibnamefont
  {Hassouneh}}, \bibinfo {author} {\bibfnamefont {S.}~\bibnamefont {Law}},
  \bibinfo {author} {\bibfnamefont {S.~F.~C.}\ \bibnamefont {Shearer}},
  \bibinfo {author} {\bibfnamefont {A.~C.}\ \bibnamefont {Brown}}, \ and\
  \bibinfo {author} {\bibfnamefont {H.~W.}\ \bibnamefont {van~der Hart}},\
  }\href {\doibase 10.1103/PhysRevA.91.031404} {\bibfield  {journal} {\bibinfo
  {journal} {Phys. Rev. A}\ }\textbf {\bibinfo {volume} {91}},\ \bibinfo
  {pages} {031404} (\bibinfo {year} {2015})}\BibitemShut {NoStop}%
\bibitem [{\citenamefont {Armstrong}\ \emph
  {et~al.}(2019{\natexlab{a}})\citenamefont {Armstrong}, \citenamefont
  {Clarke}, \citenamefont {Brown},\ and\ \citenamefont {van~der
  Hart}}]{armstrong2019}%
  \BibitemOpen
  \bibfield  {author} {\bibinfo {author} {\bibfnamefont {G.~S.~J.}\
  \bibnamefont {Armstrong}}, \bibinfo {author} {\bibfnamefont {D.~D.~A.}\
  \bibnamefont {Clarke}}, \bibinfo {author} {\bibfnamefont {A.~C.}\
  \bibnamefont {Brown}}, \ and\ \bibinfo {author} {\bibfnamefont {H.~W.}\
  \bibnamefont {van~der Hart}},\ }\href {\doibase 10.1103/PhysRevA.99.023429}
  {\bibfield  {journal} {\bibinfo  {journal} {Phys. Rev. A}\ }\textbf {\bibinfo
  {volume} {99}},\ \bibinfo {pages} {023429} (\bibinfo {year}
  {2019}{\natexlab{a}})}\BibitemShut {NoStop}%
\bibitem [{\citenamefont {van~der Hart}(1996)}]{vdh1996}%
  \BibitemOpen
  \bibfield  {author} {\bibinfo {author} {\bibfnamefont {H.~W.}\ \bibnamefont
  {van~der Hart}},\ }\href {\doibase 10.1088/0953-4075/29/14/018} {\bibfield
  {journal} {\bibinfo  {journal} {Journal of Physics B: Atomic, Molecular and
  Optical Physics}\ }\textbf {\bibinfo {volume} {29}},\ \bibinfo {pages} {3059}
  (\bibinfo {year} {1996})}\BibitemShut {NoStop}%
\bibitem [{\citenamefont {van~der Hart}(2000)}]{vdh2000}%
  \BibitemOpen
  \bibfield  {author} {\bibinfo {author} {\bibfnamefont {H.~W.}\ \bibnamefont
  {van~der Hart}},\ }\href {\doibase 10.1088/0953-4075/33/9/309} {\bibfield
  {journal} {\bibinfo  {journal} {Journal of Physics B: Atomic, Molecular and
  Optical Physics}\ }\textbf {\bibinfo {volume} {33}},\ \bibinfo {pages} {1789}
  (\bibinfo {year} {2000})}\BibitemShut {NoStop}%
\bibitem [{\citenamefont {Clementi}\ and\ \citenamefont
  {Roetti}(1974)}]{clemroe}%
  \BibitemOpen
  \bibfield  {author} {\bibinfo {author} {\bibfnamefont {E.}~\bibnamefont
  {Clementi}}\ and\ \bibinfo {author} {\bibfnamefont {C.}~\bibnamefont
  {Roetti}},\ }\href {\doibase https://doi.org/10.1016/S0092-640X(74)80016-1}
  {\bibfield  {journal} {\bibinfo  {journal} {Atomic Data and Nuclear Data
  Tables}\ }\textbf {\bibinfo {volume} {14}},\ \bibinfo {pages} {177 }
  (\bibinfo {year} {1974})}\BibitemShut {NoStop}%
\bibitem [{\citenamefont {Hutchinson}\ \emph {et~al.}(2010)\citenamefont
  {Hutchinson}, \citenamefont {Lysaght},\ and\ \citenamefont {van~der
  Hart}}]{hutchinson2010}%
  \BibitemOpen
  \bibfield  {author} {\bibinfo {author} {\bibfnamefont {S.}~\bibnamefont
  {Hutchinson}}, \bibinfo {author} {\bibfnamefont {M.~A.}\ \bibnamefont
  {Lysaght}}, \ and\ \bibinfo {author} {\bibfnamefont {H.~W.}\ \bibnamefont
  {van~der Hart}},\ }\href {\doibase 10.1088/0953-4075/43/9/095603} {\bibfield
  {journal} {\bibinfo  {journal} {Journal of Physics B: Atomic, Molecular and
  Optical Physics}\ }\textbf {\bibinfo {volume} {43}},\ \bibinfo {pages}
  {095603} (\bibinfo {year} {2010})}\BibitemShut {NoStop}%
\bibitem [{\citenamefont {Barth}\ and\ \citenamefont
  {Smirnova}(2011)}]{barth2011}%
  \BibitemOpen
  \bibfield  {author} {\bibinfo {author} {\bibfnamefont {I.}~\bibnamefont
  {Barth}}\ and\ \bibinfo {author} {\bibfnamefont {O.}~\bibnamefont
  {Smirnova}},\ }\href {\doibase 10.1103/PhysRevA.84.063415} {\bibfield
  {journal} {\bibinfo  {journal} {Phys. Rev. A}\ }\textbf {\bibinfo {volume}
  {84}},\ \bibinfo {pages} {063415} (\bibinfo {year} {2011})}\BibitemShut
  {NoStop}%
\bibitem [{\citenamefont {Kaushal}\ \emph {et~al.}(2015)\citenamefont
  {Kaushal}, \citenamefont {Morales},\ and\ \citenamefont
  {Smirnova}}]{kaushal2015}%
  \BibitemOpen
  \bibfield  {author} {\bibinfo {author} {\bibfnamefont {J.}~\bibnamefont
  {Kaushal}}, \bibinfo {author} {\bibfnamefont {F.}~\bibnamefont {Morales}}, \
  and\ \bibinfo {author} {\bibfnamefont {O.}~\bibnamefont {Smirnova}},\ }\href
  {\doibase 10.1103/PhysRevA.92.063405} {\bibfield  {journal} {\bibinfo
  {journal} {Phys. Rev. A}\ }\textbf {\bibinfo {volume} {92}},\ \bibinfo
  {pages} {063405} (\bibinfo {year} {2015})}\BibitemShut {NoStop}%
\bibitem [{\citenamefont {Kaushal}\ and\ \citenamefont
  {Smirnova}(2018)}]{kaushal2018b}%
  \BibitemOpen
  \bibfield  {author} {\bibinfo {author} {\bibfnamefont {J.}~\bibnamefont
  {Kaushal}}\ and\ \bibinfo {author} {\bibfnamefont {O.}~\bibnamefont
  {Smirnova}},\ }\href {\doibase 10.1088/1361-6455/aad132} {\bibfield
  {journal} {\bibinfo  {journal} {Journal of Physics B: Atomic, Molecular and
  Optical Physics}\ }\textbf {\bibinfo {volume} {51}},\ \bibinfo {pages}
  {174002} (\bibinfo {year} {2018})}\BibitemShut {NoStop}%
\bibitem [{\citenamefont {Liu}\ \emph {et~al.}(2018)\citenamefont {Liu},
  \citenamefont {Ni}, \citenamefont {Renziehausen}, \citenamefont {Rost},\ and\
  \citenamefont {Barth}}]{liu2018}%
  \BibitemOpen
  \bibfield  {author} {\bibinfo {author} {\bibfnamefont {K.}~\bibnamefont
  {Liu}}, \bibinfo {author} {\bibfnamefont {H.}~\bibnamefont {Ni}}, \bibinfo
  {author} {\bibfnamefont {K.}~\bibnamefont {Renziehausen}}, \bibinfo {author}
  {\bibfnamefont {J.-M.}\ \bibnamefont {Rost}}, \ and\ \bibinfo {author}
  {\bibfnamefont {I.}~\bibnamefont {Barth}},\ }\href {\doibase
  10.1103/PhysRevLett.121.203201} {\bibfield  {journal} {\bibinfo  {journal}
  {Phys. Rev. Lett.}\ }\textbf {\bibinfo {volume} {121}},\ \bibinfo {pages}
  {203201} (\bibinfo {year} {2018})}\BibitemShut {NoStop}%
\bibitem [{\citenamefont {Armstrong}\ \emph
  {et~al.}(2019{\natexlab{b}})\citenamefont {Armstrong}, \citenamefont
  {Clarke}, \citenamefont {Brown},\ and\ \citenamefont {van~der Hart}}]{pure}%
  \BibitemOpen
  \bibfield  {author} {\bibinfo {author} {\bibfnamefont {G.~S.~J.}\
  \bibnamefont {Armstrong}}, \bibinfo {author} {\bibfnamefont {D.~D.~A.}\
  \bibnamefont {Clarke}}, \bibinfo {author} {\bibfnamefont {A.~C.}\
  \bibnamefont {Brown}}, \ and\ \bibinfo {author} {\bibfnamefont {H.~W.}\
  \bibnamefont {van~der Hart}},\ }\href {https://pure.qub.ac.uk/portal}
  {\enquote {\bibinfo {title} {Pure},}\ } (\bibinfo {year}
  {2019}{\natexlab{b}})\BibitemShut {NoStop}%
\bibitem [{rep(2019)}]{repo}%
  \BibitemOpen
  \href@noop {} {}\bibinfo {howpublished} {The RMT repository
  {\url{https://gitlab.com/Uk-amor/RMT/rmt}}} (\bibinfo {year}
  {2019})\BibitemShut {NoStop}%
\end{thebibliography}%


\end{document}